\documentclass[journal]{IEEEtran}

\usepackage{amsmath,amssymb,amsfonts}
\usepackage{algorithm}
\usepackage{algpseudocode}
\usepackage{array}
\usepackage[caption=false,font=footnotesize]{subfig}
\usepackage{textcomp}
\usepackage{stfloats}
\usepackage{url}
\usepackage{graphicx}
\usepackage{cite}
\usepackage{booktabs}

\algtext*{EndIf}
\algtext*{EndFor}

\begin{document}

\title{Efficient SC-Flip Decoding for Polar Codes through Decision-Boundary Normalization}

\author{Gulinisha·Yiming,~\IEEEmembership{Student Member,~IEEE},
Zhongjun Yang,~\IEEEmembership{Student Member,~IEEE},
Zhicheng Liu,~\IEEEmembership{Member,~IEEE},
Zhiming Ma, Xumin Ni, and Yuting Liu

\thanks{Gulinisha·Yiming, Xumin Ni, and Yuting Liu are with the School of Mathematics and Statistics, Beijing Jiaotong University, Beijing 100044, China (e-mail: 25110630@bjtu.edu.cn; xmni@bjtu.edu.cn; ytliu@bjtu.edu.cn).

Zhongjun Yang is with the School of Mathematics and Information Science, Nanchang Hangkong University, Nanchang 330038, China (e-mail: yangzhj59@nchu.edu.cn).

Zhicheng Liu is with the School of Mathematics, Sichuan University,
Chengdu 610065, China (e-mail: lzc2025@scu.edu.cn).

Zhiming Ma is with the University of Chinese Academy of
Sciences, Beijing 101408, China, and also with the Academy of Mathematics
and Systems Science, CAS, Beijing 100190, China (e-mail: mazm@amt.ac.cn).

Corresponding author: Yuting Liu.}}

\maketitle

\begin{abstract}
Accurate first erroneous position (FEP) identification is key to improving the error-correction performance of successive cancellation flip (SCF) decoding. However, FEP identification remains challenging with a limited number of re-decoding attempts. To address this issue, we propose decision-boundary normalization (DBN) to improve FEP identification and develop a novel DBN-SCF decoding algorithm. DBN normalizes the magnitude of each decision log-likelihood ratio (LLR) by the standard deviation predicted by the Gaussian approximation (GA) for the corresponding bit-channel. This allows a single normalized threshold to be shared across bit-channels within each code configuration, giving it a consistent statistical meaning and avoiding per-channel calibration. DBN-SCF decoding selects candidate flip positions from the information set in the natural successive cancellation (SC) decoding order using a threshold, without sorting. Moreover, DBN-SCF decoding achieves better error-correction performance with a limited number of re-decoding attempts. It also requires lower average decoding complexity than existing SCF decoding methods at a comparable level of error-correction performance.
\end{abstract}


\section{Introduction}

\IEEEPARstart{P}{olar} codes achieve the capacity of binary-input memoryless symmetric channels under successive cancellation (SC) decoding \cite{arikan}. However, the error-correction performance of polar codes at finite codeword lengths remains limited. To improve this, SC list (SCL) decoding \cite{talvardy} and cyclic redundancy check (CRC)-aided SCL (CA-SCL) decoding \cite{niucrc} have been proposed. These methods improve decoding performance by retaining multiple candidate paths, but increase computational complexity.

SC flip (SCF) decoding \cite{afisiadis} improves SC decoding by performing additional decoding attempts with selected bit decisions reversed after a CRC failure. A key challenge in SCF decoding is identifying the first erroneous position (FEP), since correcting  the erroneous decision at the FEP can potentially prevent subsequent error propagation. The original SCF decoding selects flip candidates according to their decision log-likelihood ratio (LLR) magnitudes. Subsequent studies have developed a dynamic reliability metric \cite{dscf} , simplified implementations \cite{sdscf}, and refined candidate-selection metrics \cite{improvedscf}. More recently, thresholded SCF (Th-SCF) decoding \cite{thscf} introduced a sorting-free candidate-selection strategy based on a predetermined LLR threshold. One possible extension is to replace this common threshold with independently optimized raw-LLR thresholds for individual bit-channels. Such thresholds may improve FEP identification but require considerable per-channel calibration and parameter management. Despite these developments, FEP identification remains challenging with a limited number of re-decoding attempts, where only a few candidate positions can be tested and their selection becomes particularly important.

This motivates further investigation of the reliability information contained in decision LLRs. Under Gaussian approximation (GA) \cite{trifonov}, polarized
bit-channels have different LLR means and variances. Although
the LLR sign determines the SC decision and its magnitude measures
the distance from the decision boundary, equal magnitudes may
have different significance across bit-channels. This motivates
using bit-channel-specific statistics for FEP selection.

In this letter, we propose decision-boundary normalization (DBN) and develop DBN-SCF decoding based on it. DBN normalizes each decision-LLR magnitude by the GA-predicted standard deviation of its bit-channel, measuring its distance from the zero-decision boundary in standard-deviation units. This gives a single threshold a common statistical meaning across bit-channels. The threshold is calibrated offline for each code configuration and fixed over the considered signal-to-noise ratio (SNR) range, avoiding per-channel calibration. DBN-SCF decoding selects flip candidates by applying the normalized threshold to information positions in their natural SC order, without sorting. We also analyze FEP retention with a finite number of re-decoding attempts and evaluate decoding performance and complexity. The main contributions are summarized as follows:
\begin{enumerate}
\item We propose DBN to improve FEP identification by normalizing
the decision LLR magnitude according to the statistical scale
of each bit-channel. We also analyze how the threshold and
earlier candidates affect FEP retention with a finite number
of re-decoding attempts.

\item Based on DBN, we develop DBN-SCF decoding. Compared with existing SCF decoding methods, such as Th-SCF decoding, DBN-SCF decoding can identify the FEP more accurately and achieve better error-correction performance with lower decoding complexity.
\end{enumerate}

\section{Preliminaries}

\subsection{Polar Coding and SC Decoding}

Let $N=2^n$ be the codeword length and $K$ the code dimension, with code rate $R=K/N$. The information set is denoted by $\mathcal A\subseteq\{0,\ldots,N-1\}$ with $|\mathcal A|=K$, while the bits in the frozen set $\mathcal A^c$ are fixed to zero. A $\mathcal P(N,K)$ polar code maps a source vector $\boldsymbol u=[u_0,\ldots,u_{N-1}]$ into a codeword $\boldsymbol x=[x_0,\ldots,x_{N-1}]$ through
\begin{equation}
\boldsymbol{x}=\boldsymbol{u}G_N,
\label{eq:polar_encoding}
\end{equation}
where $G_N=F^{\otimes n}$ is the generator matrix,
$F=\begin{bmatrix}1&0\\1&1\end{bmatrix}$ denotes the kernel matrix, and
$\otimes n$ denotes the $n$-fold Kronecker product. The encoded bits are modulated using binary phase-shift keying (BPSK) and transmitted over an additive white Gaussian noise (AWGN) channel, where the received symbols are given by
\begin{equation}
y_i=1-2x_i+w_i,
\label{eq:channel_model}
\end{equation}
where $w_i\sim\mathcal N(0,\sigma_w^2)$ denotes the AWGN sample with noise variance $\sigma_w^2$.

SC decoding estimates the source bits sequentially. The source
bit and its estimate at position $i$ are denoted by $u_i$ and
$\hat u_i$, respectively. Starting from the channel LLRs
$2y_i/\sigma_w^2$, the decoder recursively computes the
decision LLRs using the $f$- and $g$-operations defined in
\cite{arikan}, where the latter depends on previously decoded
bits. Specifically, the decision LLR at position $i$ is defined as
\begin{equation}
L_i=\log\frac{P(u_i=0\mid\boldsymbol y,\hat{\boldsymbol u}_0^{i-1})}{P(u_i=1\mid\boldsymbol y,\hat{\boldsymbol u}_0^{i-1})},
\label{eq:sc_llr}
\end{equation}
where $\hat{\boldsymbol u}_0^{i-1}$ denotes the previously decoded bits. For each information position $i\in\mathcal A$, the estimated bit is determined by
\begin{equation}
\hat u_i=
\begin{cases}
0, & L_i\geq0,\\
1, & L_i<0,
\end{cases}
\label{eq:sc_decision}
\end{equation}
while frozen bits are set to zero. Since subsequent LLR calculations depend on previous decisions, an incorrect decision may affect subsequent estimates through error propagation.

\subsection{SCF Decoding and FEP Identification}

Since an incorrect SC decision may propagate to subsequent bits, correcting  the decision at the FEP can potentially eliminate the resulting errors. SCF decoding \cite{afisiadis} exploits this property by performing additional SC decoding attempts with selected bit decisions reversed. Its procedure consists of three steps:

\begin{enumerate}
\item \textit{Initial decoding:} Perform SC decoding and verify the decoded sequence using a CRC function. If the CRC check passes, decoding terminates.
\item \textit{Candidate selection:} If the CRC check fails, construct an ordered list of candidate flip positions $\mathcal F=(f_1,\ldots,f_T)$, where $T\leq T_{\max}$ and $T_{\max}$ represents the maximum number of additional SC decoding attempts.
\item \textit{Flip and re-decoding:} In each re-decoding attempt, the SC decisions before $f_t$ are reused, the decision at $f_t$ is flipped, and subsequent bits are re-estimated through the SC recursion. Decoding terminates when the CRC passes or all candidates have been tested.
\end{enumerate}

The main challenge in SCF decoding lies in Step 2, where the candidate list should contain the FEP. When the SC decoding result is erroneous, the FEP is defined as
\begin{equation}
i_{\rm F}=\min\{i\in\mathcal A:\hat u_i\neq u_i\}.
\label{eq:fep}
\end{equation}
Note that all decisions preceding $i_{\rm F}$ are correct, while subsequent errors may result from error propagation. Since the real FEP is unknown to the decoder, including it in the candidate list remains challenging, particularly when the number of re-decoding attempts $T_{\max}$ is small.

Addressing this problem, several efficient FEP identification methods have been developed. Min-LLR SCF decoding \cite{afisiadis} uses the decision LLR magnitude, i.e., $|L_i|$, as the reliability measure, where smaller magnitudes indicate less reliable decisions. Th-SCF decoding \cite{thscf} further introduces a predefined threshold on the decision LLR magnitude to characterize unreliable decisions. D-SCF decoding \cite{dscf} employs a dynamic reliability metric  that accounts for current reliability and preceding decisions.

Although these methods employ different candidate-selection strategies, FEP identification remains challenging with a limited number of re-decoding attempts, e.g., $T_{\max}=2$. Moreover, the existing FEP identification criteria primarily rely on observed decision LLRs or derived reliability metrics, without explicitly normalizing the decision-boundary distance by the statistical scale of each polarized bit-channel. This motivates further investigation into whether bit-channel-specific LLR statistics can improve FEP identification.

\section{Decision-Boundary Normalization for FEP Identification}

This section investigates the statistical characteristics of decision LLRs across polarized bit-channels to motivate GA-guided FEP identification. We then define a normalized decision-boundary distance, derive the corresponding error-coverage probability, and evaluate its effectiveness in FEP identification.

\subsection{Gaussian Statistics of Decision LLRs}

For simplicity of analysis, an all-zero transmitted codeword is assumed. According to the GA \cite{trifonov}, under correct preceding decisions, the decision LLR of the $i$-th polarized bit-channel can be approximated as
\begin{equation}
L_i\sim\mathcal N\left(\mu_i(\gamma),2\mu_i(\gamma)\right),
\label{eq:ga_distribution}
\end{equation}
where $\mu_i(\gamma)$ denotes the GA-predicted LLR mean at operating SNR $\gamma$, and $\sigma_i(\gamma)=\sqrt{2\mu_i(\gamma)}$ is the corresponding standard deviation. Due to channel polarization, different bit-channels generally exhibit different LLR statistical scales. Normalizing $L_i$ by its standard deviation yields
\begin{equation}
\frac{L_i}{\sigma_i(\gamma)}
\sim\mathcal N\left(
\frac{\mu_i(\gamma)}{\sigma_i(\gamma)},1
\right),
\label{eq:standardized_llr}
\end{equation}
indicating that the standardized LLRs have unit variance but different mean distances from the zero-decision boundary.

Under the same GA assumptions, an erroneous decision occurs
when $L_i<0$. The conditional bit-channel error probability
is defined as $P_{e,i}^{\rm GA}(\gamma)=\Pr(L_i<0)$ and is
given by
\begin{equation}
\begin{aligned}
P_{e,i}^{\rm GA}(\gamma)
&=\Pr\left(\frac{L_i}{\sigma_i(\gamma)}<0\right)\\
&=Q\left(\frac{\mu_i(\gamma)}{\sigma_i(\gamma)}\right)\\
&=Q\left(\sqrt{\frac{\mu_i(\gamma)}{2}}\right).
\end{aligned}
\label{eq:ga_error}
\end{equation}
where $Q(x)=\frac{1}{\sqrt{2\pi}}\int_x^{\infty}
\exp(-t^2/2)\,\mathrm{d}t$ denotes the standard Gaussian tail
function. Since $\sigma_i(\gamma)=\sqrt{2\mu_i(\gamma)}$,
the normalized mean distance is
$\mu_i(\gamma)/\sigma_i(\gamma)=\sqrt{\mu_i(\gamma)/2}$.
According to \eqref{eq:ga_error}, a larger normalized mean
distance corresponds to a lower bit-channel error probability.
This motivates normalizing the observed decision LLR magnitude
by the corresponding standard deviation to account for
differences in LLR scales across bit-channels.

\subsection{Normalized Distance and Error Coverage}

To account for differences in bit-channel scales, we define
the DBN-based distance as
\begin{equation}
\begin{aligned}
d_i(\gamma)
&=\frac{|L_i|}{\sigma_i(\gamma)}\\
&=\frac{|L_i|}{\sqrt{2\mu_i(\gamma)}}.
\end{aligned}
\label{eq:normalized_distance}
\end{equation}
A common raw-LLR threshold treats equal magnitudes identically, although their significance depends on the statistical scale of the corresponding bit-channel. DBN measures $|L_i|$ in units of $\sigma_i(\gamma)$, the GA-predicted standard deviation, making candidate screening more comparable across bit-channels. The mean is not subtracted because FEP identification concerns the distance from the SC decision boundary $L_i=0$, rather than the deviation from the GA mean $\mu_i(\gamma)$.

To determine whether erroneous decisions satisfy the threshold, let $S_i=(-1)^{u_i}L_i$ denote the sign-corrected decision LLR. Under the all-zero codeword assumption, $S_i=L_i$. The sign correction accounts for channel symmetry. Under the GA model, define
\begin{equation}
Y_i=\frac{S_i}{\sigma_i(\gamma)},
\quad
a_i(\gamma)=\sqrt{\frac{\mu_i(\gamma)}{2}},
\label{eq:sign_corrected_llr}
\end{equation}
such that $Y_i\sim\mathcal N(a_i(\gamma),1)$ approximately. An erroneous decision corresponds to $Y_i<0$, while the normalized screening condition with threshold $\theta\geq0$ is equivalent to $|Y_i|\leq\theta$.

Under the GA model, the probability that an erroneous decision satisfies the threshold is
\begin{equation}
\begin{aligned}
F_i^{\rm err}(\theta)
&=\Pr(|Y_i|\leq\theta\mid Y_i<0)\\
&=\frac{\Pr(-\theta\leq Y_i<0)}
{\Pr(Y_i<0)}\\
&=1-\frac{Q(a_i(\gamma)+\theta)}
{Q(a_i(\gamma))}.
\end{aligned}
\label{eq:ga_error_coverage}
\end{equation}
Its derivative with respect to $\theta$ is
\begin{equation}
\frac{\partial F_i^{\rm err}(\theta)}
{\partial\theta}
=
\frac{\phi(a_i(\gamma)+\theta)}
{Q(a_i(\gamma))}>0,
\label{eq:coverage_derivative}
\end{equation}
where $\phi(\cdot)$ denotes the standard Gaussian probability density function. Thus, the threshold coverage of erroneous decisions increases
with $\theta$. However, this conditional coverage is not the
actual FEP identification rate, which also depends on
preceding SC decisions and the number of re-decoding attempts.

\begin{figure}[!t]
\centering
\includegraphics[width=0.92\columnwidth]{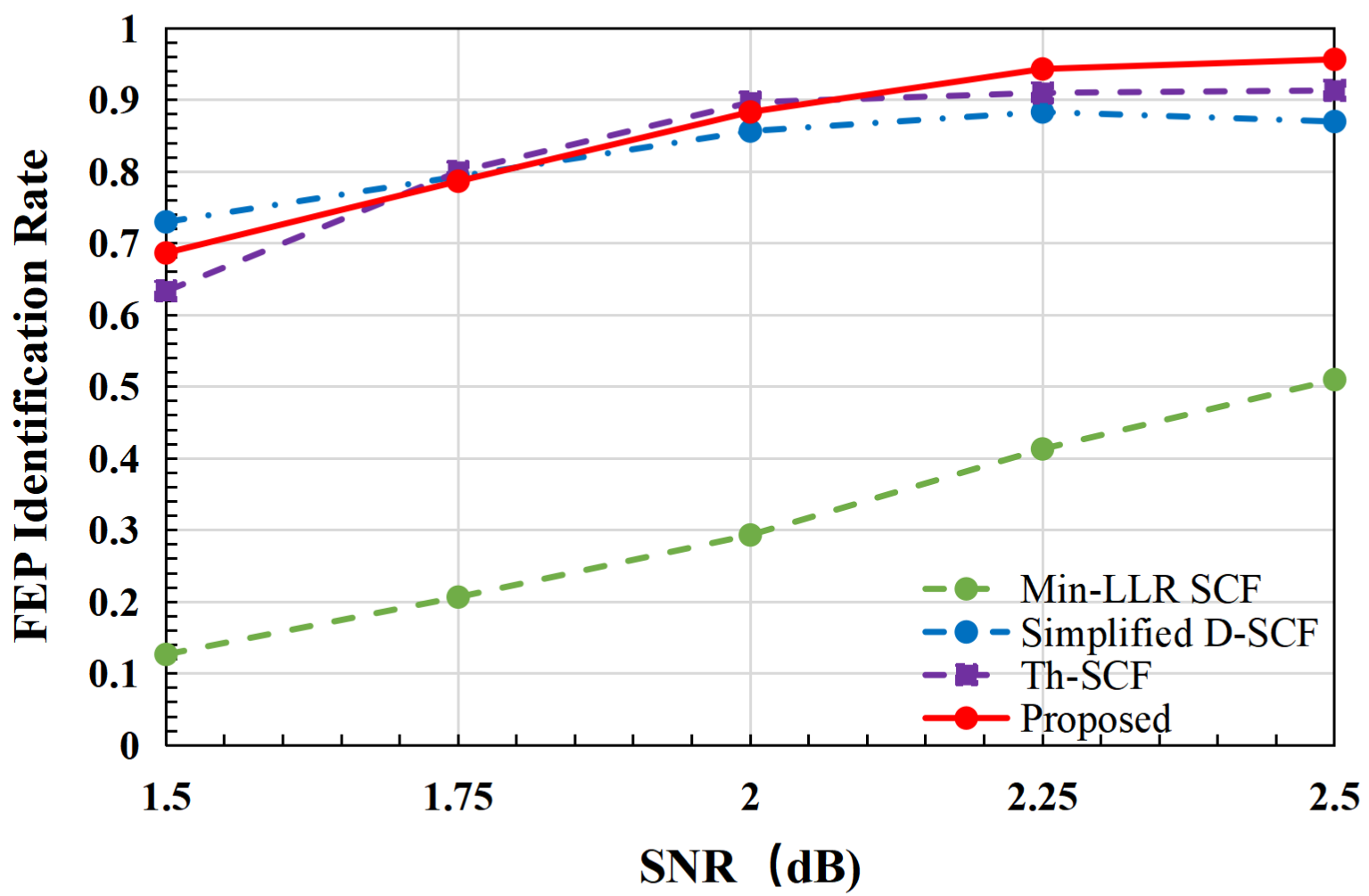}
\caption{FEP identification rates of different candidate-selection strategies versus SNR under $T_{\max}=2$.}
\label{fig:pfind}
\end{figure}

Fig.~\ref{fig:pfind} compares the FEP identification rates of different candidate-selection strategies under $T_{\max}=2$. The FEP identification rate is defined as the ratio of the frames where the real FEP is included in the candidate list to the total number of erroneous SC decoding frames. The proposed normalized screening strategy achieves higher FEP identification rates than existing methods at higher operating SNRs, improving SCF error-correction performance with fewer re-decoding attempts.

\section{Decision-Boundary Normalized SCF Decoding}

This section proposes DBN-SCF decoding, which is based on the DBN distance defined in Section III. We first describe the construction of the FEP set and characterize FEP retention with a finite number of re-decoding attempts. The complete decoding procedure is then presented.

\subsection{Efficient Identification of FEP}

The objective of candidate selection is to construct the FEP set that is likely to contain the real FEP with a limited number of re-decoding attempts. Based on the normalized distance in \eqref{eq:normalized_distance}, an information position $i$ is considered for selection if $d_i(\gamma)\leq\theta$, where $\theta\geq0$ is the normalized threshold. Equivalently, the screening condition in the raw-LLR domain is
\begin{equation}
|L_i|\leq T_i^{\rm eff}(\gamma),
\quad
T_i^{\rm eff}(\gamma)
=\theta\sqrt{2\mu_i(\gamma)}.
\label{eq:adaptive_threshold}
\end{equation}
Different from Th-SCF decoding \cite{thscf}, which employs a common raw-LLR threshold, DBN-SCF decoding induces bit-channel- and SNR-dependent effective thresholds while retaining a sorting-free selection procedure.

For each code configuration, the threshold $\theta$ is calibrated offline through extensive Monte Carlo simulations by comparing block error rate (BLER) performance over the considered SNR range. The value yielding the best overall BLER performance is selected and fixed for online decoding. A common threshold is used across the considered SNR range for each code configuration, while different codeword lengths and rates may use different threshold values. The GA parameters $\{\mu_i(\gamma)\}$ are also precomputed for the corresponding code configuration and operating SNR. Therefore, neither threshold search nor GA construction is required during online candidate selection.

Let
\begin{equation}
\mathcal E_\theta
=\{i\in\mathcal A:d_i(\gamma)\leq\theta\}
\label{eq:eligible_set}
\end{equation}
denote the set of positions that satisfy the threshold. The FEP set is constructed by considering the information positions in their natural SC order and retaining the first $T_{\max}$ positions that satisfy the threshold, or all such positions if fewer than $T_{\max}$ are found. This procedure has $\mathcal{O}(K)$ candidate-selection complexity without sorting. The construction procedure is summarized in Algorithm~\ref{alg:construct_dbn}.

\begin{algorithm}[!t]
\caption{Constructing the FEP Set}
\label{alg:construct_dbn}
\begin{algorithmic}[1]
\Require $\mathcal A$, $\{L_i\}$, $\{\mu_i(\gamma)\}$, $\theta$, $T_{\max}$
\Ensure FEP set $\mathcal F_D$
\State $\mathcal F_D\gets\emptyset$
\For{$i\in\mathcal A$ in SC order}
    \State $d_i\gets |L_i|/\sqrt{2\mu_i(\gamma)}$
    \If{$d_i\leq\theta$ \textbf{and} $|\mathcal F_D|<T_{\max}$}
        \State Append $i$ to $\mathcal F_D$
    \EndIf
\EndFor
\State \textbf{return} $\mathcal F_D$
\end{algorithmic}
\end{algorithm}

To analyze FEP retention with a finite number of re-decoding
attempts, let $\mathcal F_D^{(T)}$ denote the first $T$ positions
in SC order that satisfy the threshold, or all such positions
if fewer than $T$ exist. Let $N_{<{\rm F}}(\theta)$ denote the
number of information positions before the real FEP that
satisfy the threshold. It is calculated as
\begin{equation}
N_{<{\rm F}}(\theta)
=\sum_{\substack{j\in\mathcal A\\j<i_{\rm F}}}
\mathbb{I}\{d_j(\gamma)\leq\theta\}.
\label{eq:earlier_candidates}
\end{equation}
Note that $\mathbb{I}\{\cdot\}$ denotes an indicator function
that equals one if the condition holds and zero otherwise.

The real FEP is retained in the candidate list if and only if
\begin{equation}
\begin{aligned}
i_{\rm F}\in\mathcal F_D^{(T)}
\iff {}&
[d_{i_{\rm F}}(\gamma)\leq\theta]\land[N_{<{\rm F}}(\theta)<T].
\end{aligned}
\label{eq:fep_retention}
\end{equation}

Let $\mathcal E_{\rm SC}$ denote the event of an erroneous initial SC decoding. The FEP threshold coverage is defined as
\begin{equation}
C_{\rm F}(\theta)
=\Pr\left(d_{i_{\rm F}}(\gamma)\leq\theta
\mid\mathcal E_{\rm SC}\right).
\label{eq:fep_coverage}
\end{equation}
Different from the conditional GA error coverage in \eqref{eq:ga_error_coverage}, $C_{\rm F}(\theta)$ characterizes the actual FEP distribution. From \eqref{eq:fep_retention}, the FEP identification rate $P_{\rm Find}(\theta,T)=\Pr(i_{\rm F}\in\mathcal F_D^{(T)}\mid\mathcal E_{\rm SC})$ satisfies the exact decomposition
\begin{equation}
\begin{aligned}
P_{\rm Find}(\theta,T)
&=C_{\rm F}(\theta)-P_{\rm excl}(\theta,T),\\
P_{\rm excl}(\theta,T)
&=\Pr\!\left(d_{i_{\rm F}}(\gamma)\leq\theta,\,
N_{<{\rm F}}(\theta)\geq T
\mid\mathcal E_{\rm SC}\right).
\end{aligned}
\label{eq:pfind_decomposition}
\end{equation}
Note that $P_{\rm excl}(\theta,T)$ denotes the probability that the real FEP satisfies the threshold but is excluded because at least $T$ preceding positions also satisfy the threshold. Increasing $\theta$ can improve threshold coverage but may also increase this exclusion probability. Thus, $\theta$ must balance threshold coverage against
exclusion by earlier candidates when $T$ is small.

\subsection{The DBN-SCF Decoding}

Based on the FEP set constructed in Sec. IV-A, DBN-SCF
decoding performs re-decoding attempts to correct erroneous
bit decisions. Let $\mathrm{CRCCheck}(\cdot)$ denote the CRC verification function, which returns true if decoding is successful and
false otherwise.

The DBN-SCF decoding first performs SC decoding on the
received sequence $\boldsymbol y$ to obtain the estimated source vector
$\hat{\boldsymbol u}^{(0)}$ and the decision LLRs $\boldsymbol L$.
If $\mathrm{CRCCheck}(\hat{\boldsymbol u}^{(0)})=\mathrm{true}$,
decoding terminates. Otherwise, the FEP set $\mathcal F_D$ is constructed
using Algorithm 1. For each candidate $f_t\in\mathcal F_D$, SC decoding
is restarted from the original received sequence, with the
decision at position $f_t$ forcibly reversed and all subsequent
bits re-estimated through the SC recursion. Each attempt flips
only one position and is independent of previous attempts.
Decoding terminates when $\mathrm{CRCCheck}(\hat{\boldsymbol u})=\mathrm{true}$
or all candidates have been tested. Note that the GA parameters
$\{\mu_i(\gamma)\}$ can be computed offline, so no additional online
complexity is required to compute these parameters for FEP
identification. The proposed DBN-SCF decoding is summarized in
Algorithm 2.

\begin{algorithm}[!t]
\caption{The DBN-SCF Decoding}
\label{alg:dbnscf}
\begin{algorithmic}[1]
\Require $\boldsymbol y$, $\mathcal A$, $\{\mu_i(\gamma)\}$, $\theta$, $T_{\max}$
\Ensure $\hat{\boldsymbol u}$, CRC status
\State $(\hat{\boldsymbol u}^{(0)},\boldsymbol L)
\gets\mathrm{SC}(\boldsymbol y)$
\If{$\mathrm{CRCCheck}(\hat{\boldsymbol u}^{(0)})
=\mathrm{true}$}
    \State \textbf{return} $(\hat{\boldsymbol u}^{(0)},\mathrm{true})$
\EndIf
\State $\mathcal F_D\gets
\mathrm{ConstructDBN}(\mathcal A,\boldsymbol L,
\{\mu_i(\gamma)\},\theta,T_{\max})$
\For{$f_t\in\mathcal F_D$}
    \State $\hat{\boldsymbol u}\gets
    \mathrm{SC}(\boldsymbol y,\mathrm{flip}=f_t)$
    \If{$\mathrm{CRCCheck}(\hat{\boldsymbol u})=\mathrm{true}$}
        \State \textbf{return} $(\hat{\boldsymbol u},\mathrm{true})$
    \EndIf
\EndFor
\State \textbf{return} $(\hat{\boldsymbol u}^{(0)},\mathrm{false})$
\end{algorithmic}
\end{algorithm}

\section{Simulation Results}

This section evaluates DBN-SCF decoding using a 16-bit CRC specified in 5G NR and polar codes constructed by GA at a $\mathit{design}$-SNR of $2.35$ dB. We compare DBN-SCF decoding with SC, Min-LLR SCF\cite{afisiadis}, Simplified D-SCF\cite{sdscf}, Th-SCF \cite{thscf}, and CA-SCL decoding \cite{niucrc} with list size of two, denoted as CA-SCL(2). We set $T_{\max}=2$ to evaluate DBN-SCF decoding with a limited number of re-decoding attempts. Specifically, $\theta=\{0.90,0.75,0.85\}$ for $N=\{128,1024,4096\}$ at $R=1/2$, and $\theta=\{1.00,0.85\}$ for $R=\{1/4,3/4\}$ at $N=4096$.

\subsection{BLER Performance}

\begin{figure}[!t]
\centering
\subfloat[]{
    \includegraphics[width=0.49\columnwidth]{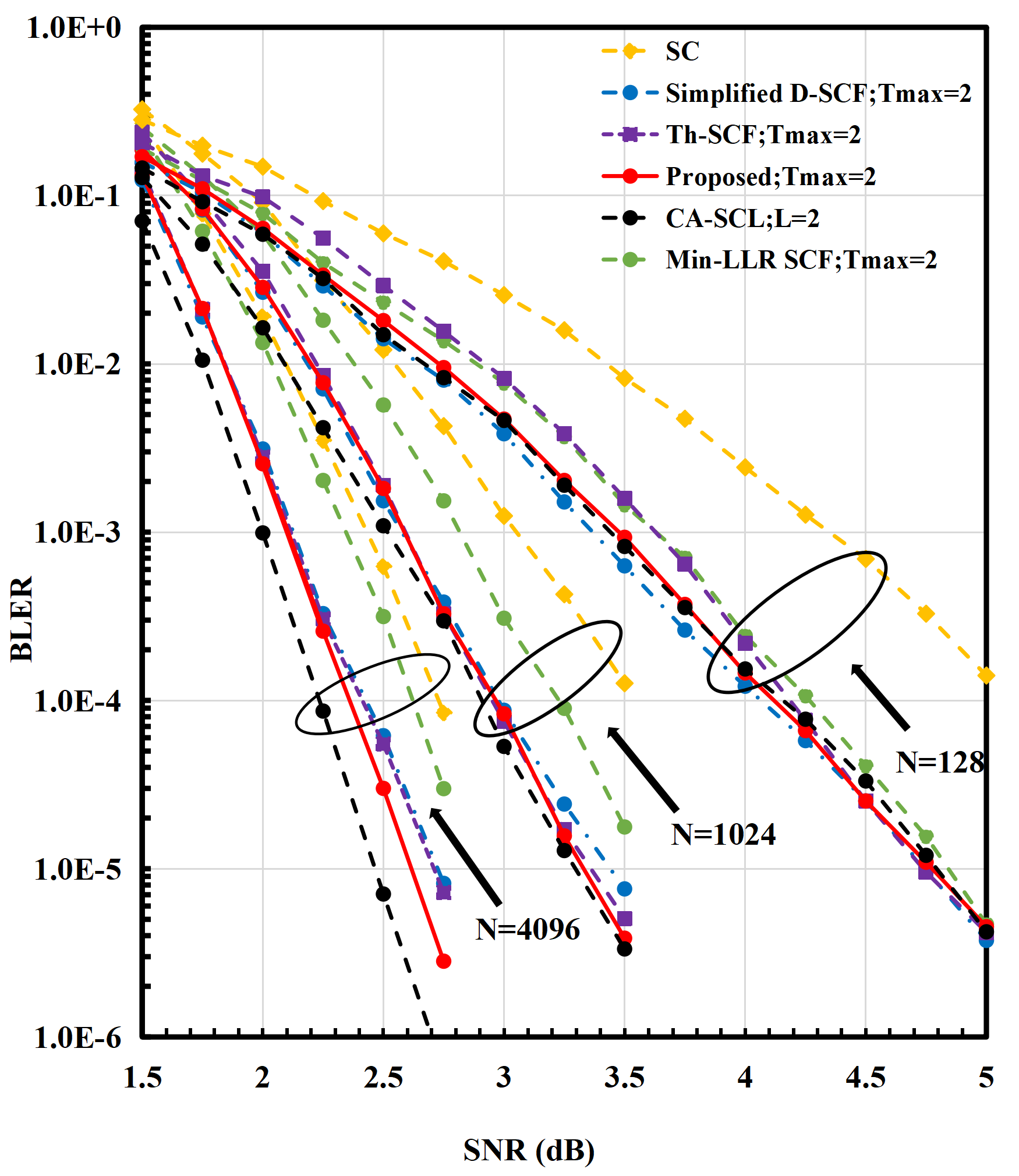}
    \label{fig:bler_length}
}%
\hspace{-0.05\columnwidth}
\subfloat[]{
    \includegraphics[width=0.49\columnwidth]{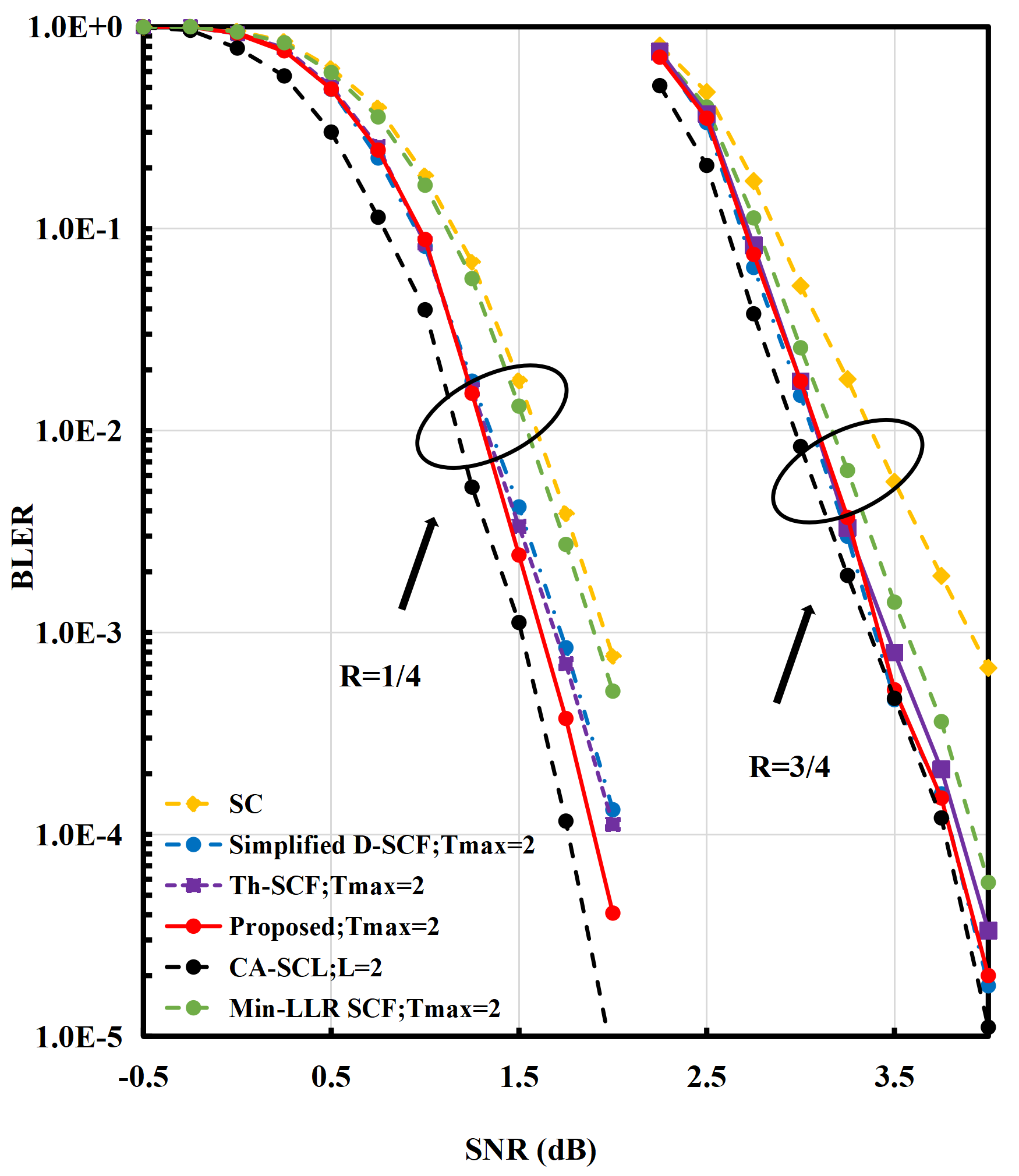}
    \label{fig:bler_rate}
}%
\caption{BLER performance for various codeword lengths and rates, where $T_{\max}=2$. (a) $N\in\{128,1024,4096\}$ at $R=1/2$; (b) $R\in\{1/4,3/4\}$ at $N=4096$.}
\label{fig:bler_robust}
\end{figure}

Fig.~\ref{fig:bler_robust}(a) shows the BLER performance of DBN-SCF decoding for different codeword lengths at $R=1/2$. With only two re-decoding attempts ($T_{\max}=2$), DBN-SCF decoding achieves a lower BLER than Min-LLR SCF decoding and Th-SCF decoding for $N=128$, while Simplified D-SCF decoding achieves a slightly lower BLER. At this short codeword length, channel polarization is less pronounced, resulting in smaller differences in bit-channel LLR scales and a more limited gain from DBN normalization. For $N=1024$ and $4096$, DBN-SCF decoding achieves more pronounced BLER gains, particularly at high SNRs.

Fig.~\ref{fig:bler_robust}(b) shows the BLER performance for different code rates at $N=4096$. DBN-SCF decoding maintains competitive performance for both low-rate and high-rate polar codes, demonstrating the applicability of the normalized screening strategy across different code configurations.

\subsection{Threshold Coverage Analysis}

To validate the offline-selected threshold, a genie-aided SC
\cite{liuling} experiment is conducted, in which preceding bit
decisions are supplied correctly. Let $\widetilde u_i^{\rm G}$
and $L_i^{\rm G}$ denote the tentative bit decision before
genie correction and its decision LLR, respectively. The
genie-aided FEP and its normalized distance are defined as
$i_{\rm G}=\min\{i\in\mathcal A:\widetilde u_i^{\rm G}\neq u_i\}$
and $Z_{\rm G}=|L_{i_{\rm G}}^{\rm G}|/
\sqrt{2\mu_{i_{\rm G}}(\gamma)}$, respectively.

For $N=4096$ and $R=1/2$, the selected threshold
$\theta=0.85$ covers approximately $91\%$ of genie-aided
FEPs.

\subsection{Average Decoding Complexity}
The complexity of DBN-SCF decoding is determined by the costs of SC decoding and FEP set construction. Let $T_{\rm avg}$ denote
the average number of SC decoding attempts per received frame.
Since each attempt has complexity $\mathcal{O}(N\log_2N)$ and
Algorithm~\ref{alg:construct_dbn} constructs the FEP set with
$\mathcal{O}(K)$ complexity, the average complexity of DBN-SCF decoding is
\begin{equation}
\Phi = T_{\rm avg}N\log_2N+K.
\label{eq:dbn_complexity}
\end{equation}
Note that $K$ is a lower-order term and can be neglected for large $N$.

\begin{figure}[!t]
\centering
\includegraphics[width=0.92\columnwidth]{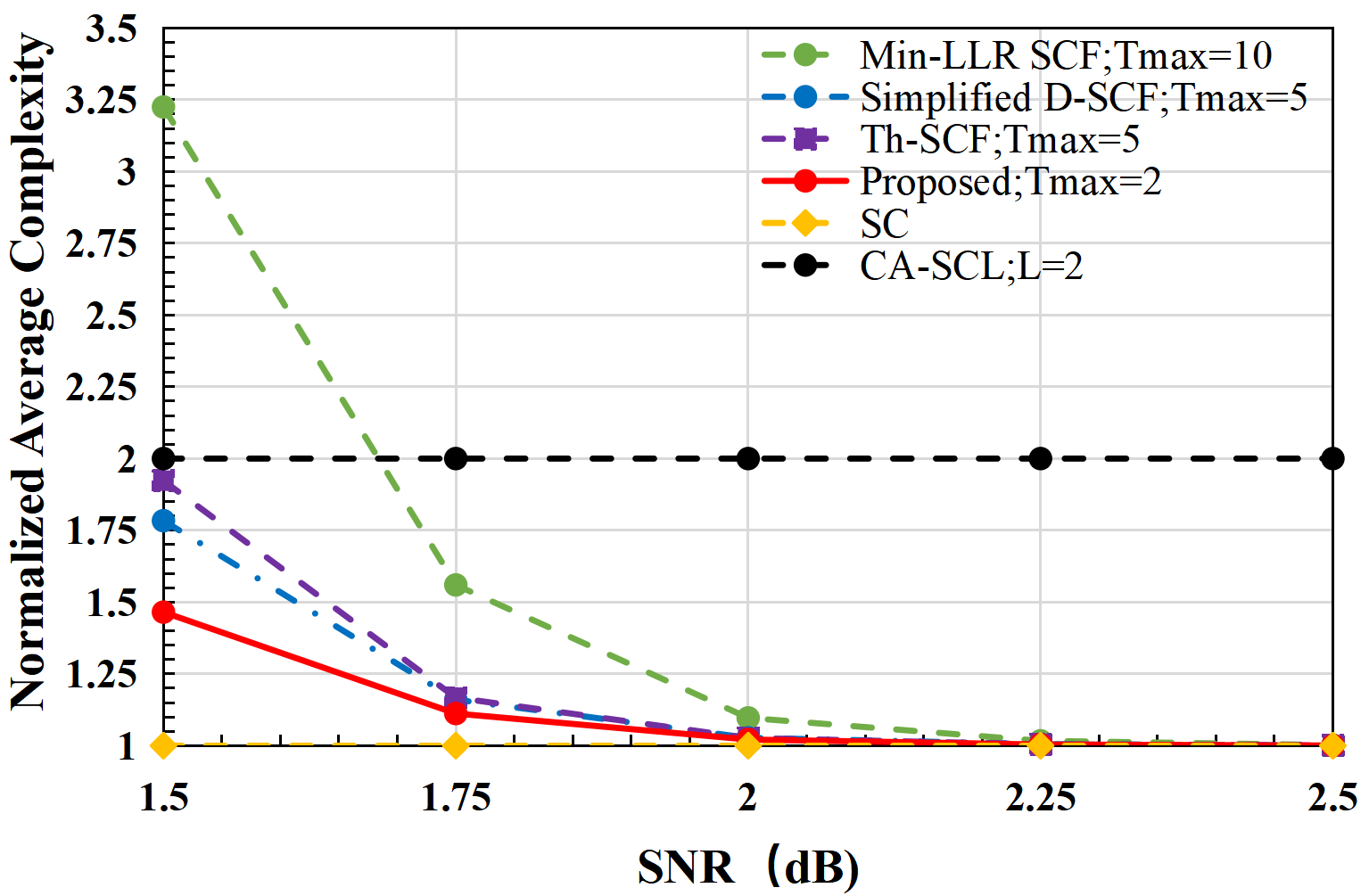}
\caption{Average normalized decoding complexity for a polar code with $N=4096$, $R=1/2$, and CRC16.}
\label{fig:avg_complexity}
\end{figure}
Fig.~\ref{fig:avg_complexity} compares the average normalized decoding complexity of different algorithms, where the normalization factor is the computational complexity of one SC decoding attempt. To compare decoding complexity at similar error-correction performance, the numbers of re-decoding attempts are selected to achieve a BLER of approximately $10^{-4}$. As shown in the figure, DBN-SCF decoding requires fewer SC-equivalent decoding attempts than existing SCF decoding methods. Compared with CA-SCL(2), DBN-SCF decoding also exhibits lower average normalized decoding complexity while maintaining competitive BLER performance.

\section{Conclusion}

In this letter, we propose DBN-SCF decoding and analyze FEP retention with a finite number of re-decoding attempts. Simulation results show that DBN-SCF decoding improves error-correction performance with only two re-decoding attempts and requires lower average decoding complexity than existing SCF methods at a comparable BLER. For a fixed threshold, the FEP identification rate approaches $C_{\rm F}(\theta)$ as $T_{\max}$ increases. Extending FEP coverage by combining DBN with methods such as Th-SCF is left for future work.


\begin{thebibliography}{11}

\bibitem{arikan}
E. Ar{\i}kan, ``Channel polarization: A method for constructing capacity-achieving codes for symmetric binary-input memoryless channels,'' \emph{IEEE Trans. Inf. Theory}, vol. 55, no. 7, pp. 3051--3073, Jul. 2009.

\bibitem{talvardy}
I. Tal and A. Vardy, ``List decoding of polar codes,'' \emph{IEEE Trans. Inf. Theory}, vol. 61, no. 5, pp. 2213--2226, May 2015.

\bibitem{niucrc}
K. Niu and K. Chen, ``CRC-aided decoding of polar codes,'' \emph{IEEE Commun. Lett.}, vol. 16, no. 10, pp. 1668--1671, Oct. 2012.

\bibitem{afisiadis}
O. Afisiadis, A. Balatsoukas-Stimming, and A. Burg, ``A low-complexity improved successive cancellation decoder for polar codes,'' in \emph{Proc. Asilomar Conf. Signals, Syst., Comput.}, Nov. 2014, pp. 2116--2120.

\bibitem{dscf}
L. Chandesris, V. Savin, and D. Declercq, ``Dynamic-SCFlip decoding of polar codes,'' \emph{IEEE Trans. Commun.}, vol. 66, no. 6, pp. 2333--2345, Jun. 2018.

\bibitem{sdscf}
F. Ercan, T. Tonnellier, N. Doan, and W. J. Gross, ``Simplified dynamic SC-flip polar decoding,'' in \emph{Proc. IEEE ICASSP}, May 2020, pp. 1733--1737.

\bibitem{improvedscf}
F. Ercan, C. Condo, and W. J. Gross, ``Improved bit-flipping algorithm for successive cancellation decoding of polar codes,'' \emph{IEEE Trans. Commun.}, vol. 67, no. 1, pp. 61--72, Jan. 2019.


\bibitem{thscf}
Z. Liu, L. Yao, S. Yuan, G. Yan, Z. Ma, and Y. Liu, ``Threshold successive cancellation flip decoding algorithm for polar codes: Design and performance,'' \emph{Entropy}, vol. 27, no. 6, Art. no. 626, Jun. 2025.

\bibitem{trifonov}
P. Trifonov, ``Efficient design and decoding of polar codes,'' \emph{IEEE Trans. Commun.}, vol. 60, no. 11, pp. 3221--3227, Nov. 2012.

\bibitem{liuling}
L. Liu, Q. Cao, L. Li, and B. Bai, ``On polar coding with feedback,'' arXiv:2601.09222, 2026.

\end{thebibliography}
\end{document}